\documentclass[aps,prb,twocolumn,showpacs]{revtex4}
\usepackage{graphicx}
\usepackage{amssymb}

\newcommand{\TMO}{TbMnO$_3$}
\newcommand{\GMO}{GdMnO$_3$}

\newcommand{\PMO}{PrMnO$_3$}
\newcommand{\NMO}{NdMnO$_3$}
\newcommand{\HMO}{HoMnO$_3$}
\newcommand{\YMO}{YMnO$_3$}

\newcommand{\EMO}{EuMnO$_3$}

\newcommand{\TbGd}{Tb$_{1-x}$Gd$_x$MnO$_3$}
\newcommand{\LaGd}{La$_{1-y}$Gd$_y$MnO$_3$}
\newcommand{\EuY}{Eu$_{1-x}$Y$_x$MnO$_3$}

\newcommand{\etal}{\textit{et al.}}

\begin{document}


\title{The multiferroic phases of Eu$_{1-x}$Y$_x$MnO$_3$}

\author{J. Hemberger$^1$, F.~Schrettle$^1$, A. Pimenov$^1$, P. Lunkenheimer$^1$, V.Yu.~Ivanov$^2$, A.A.~Mukhin$^2$,
A.M.~Balbashov$^3$, and A. Loidl$^1$}
\address{%
$^1$Experimentalphysik V, Center for Electronic Correlations and Magnetism, \\
University of Augsburg, D-86135 Augsburg, Germany \\ %
$^2$ General Physics Institute of the Russian Academy of Sciences,
38 Vavilov Street, 119991 Moscow, Russia \\ %
$^3$ Moscow Power Engineering Institute, 14 Krasnokasarmennaja Street, 111250 Moscow, Russia %
}

\begin{abstract}
We report on structural, magnetic, dielectric, and thermodynamic
properties of \EuY\ for Y doping levels $0 \leq x < 1$. This
system resembles the multiferroic perovskite manganites $R$MnO$_3$
(with $R=$~Gd, Dy, Tb) but without the interference of magnetic
contributions of the $4f$-ions. In addition, it offers the
possibility to continuously tune the influence of the $A$-site
ionic radii. For small concentrations $x\leq 0.1$ we find a canted
antiferromagnetic and paraelectric groundstate. For higher
concentrations $x\geq 0.3$ ferroelectric polarization coexists
with the features of a long wavelength incommensurate spiral
magnetic phase analogous to the observations in \TMO. In the
intermediate concentration range around $x\approx 0.2$ a
multiferroic scenario is realized combining weak ferroelectricity
and weak ferromagnetism, presumably due to a canted spiral
magnetic structure.
\end{abstract}

\pacs{71.45.Gm, 75.30.-m, 75.30.Kz, 77.84.Bw, 77.90.+k}  


\maketitle

\section{Introduction}

In recent years multiferroics have attracted an increasing
scientific and technological interest in the large community
working on functional transition-metal compounds.\cite{fiebig:05}
Within this rare class of materials magnetic order coexists with
long range polar order and both order-parameters are strongly
coupled. Prominent examples for such multiferroics may be found
among Cr-based spinels \cite{hemberger:05nat,fichtl:05prb}, the
kagom\'e staircase compound Ni$_3$V$_2$O$_8$ \cite{lawes:05}, rare
earth manganites like hexagonal \YMO\ and \HMO\
\cite{lottermoser:04,fiebig:02}, orthorombic
TbMn$_2$O$_5$\cite{hur:04}, or finally perovskites like \TMO.
\cite{kimura:03}  

In this latter class of heavy rare earth compounds $R$MnO$_3$
($R=$~Gd, Tb, Dy) \cite{kimura:05,goto:04} finite ferroelectric
polarization is induced due to the partial frustration and
corresponding long range modulation of the magnetic structure. The
importance of the Dzyaloshinskii-Moriya interaction for the
occurrence of ferroelectricity in non-collinear magnets has been
pointed out recently. \cite{katsura:05,ederer:06,sergienko:06}
Starting from the lanthanide manganite with the perovskite
tolerance factor close to unity, the substitution of La at the
$A$-sites by rare-earth elements with smaller ionic radii (from Pr
to Ho) leads to a successive increase of the orthorombic
distortion, accompanied by a decrease of the Mn-O-Mn bond angles
and an increase of the buckling and tilting angles of the
MnO$_6$-octahedra, respectively.\cite{goto:04} Using
high-resolution x-ray diffraction and the refinement of the oxygen
positions, this angle has recently been determined for the
complete series of pure rare-earth manganites. \cite{kimura:03}
Equivalent results can be obtained when continuously replacing
La$^{3+}$ by smaller ions like in \LaGd.\cite{hemberger:04prb} The
enhanced tilting of the MnO$_6$-octahedra leads to an increasing
importance of antiferromagnetic (AFM) next-nearest neighbor
interactions competing with the nearest neighbor superexchange
within the ferromagnetic (FM) $ab$-planes of the $A$-type AFM
structure. This weakening of the effective magnetic interaction
promotes a tendency towards frustration and complex
spin-states.\cite{kimura:03prb} Hence, the transition temperature
into the $A$-type AFM phase is not only reduced, but for the heavy
rare earth compounds with smaller ionic radii an incommensurate
magnetic structure is established for temperatures below
$T_N\approx50$~K.\cite{hemberger:04prb,kimura:03prb}
Within these incommensurate magnetic phases, 
magnetoelectric coupling via long-range modulation of the magnetic
structure leads to the loss of inversion symmetry and the onset of
ferroelectric polarization. \cite{mostovoy:06} The onset of
ferroelectricity is connected with the change of the magnetic
structure from a sinusoidal to a helicoidal
modulation.\cite{kenzelmann:05,mostovoy:06} Very recently
excitations of the multiferroic state have been observed
experimentally. \cite{pimenov:06} These new collective modes,
which are responsible for ferroelectricity, but are of magnetic
origin, have also been described theoretically. \cite{katsura:06}
The existence of a ferroelectric lattice distortion is especially
remarkable due to the fact that these compounds possess a quite
robust Jahn-Teller (JT) type orbital order, which sets in at
temperatures well above 1000~K. From a principal point of view
Jahn-Teller active orbital degrees of freedom are the "natural
enemy" of off-center ferroelectric distortions.\cite{hill:02}
Undistorted orbital electron-density distributions are
point-symmetric carrying no electric dipole moment. Hence, if the
lattice is able to relax into a lower symmetry via a Jahn-Teller
distortion this transition is expected not to be dipolar. In this
sense the ferroelectricity in the heavy rare earth manganites
exists not due to, but despite the superimposed orbital order and
comparing the relevant energy scales it has to be regarded as a
second order effect. Therefore the ferroelectric distortion is
rather weak and so far could not be proven directly using
high-resolution neutron or x-ray diffraction techniques. However,
the onset of ferroelectricity was documented via pyroelectric
measurements and is connected to a distinct anomaly in the real
part of the dielectric permittivity \cite{kimura:05} and clear
features in the anisotropic thermal expansion.\cite{baier:06} This
underlines that the ferroelectric distortions are driven by
frustrated magnetic interactions via local exchange-striction.
\cite{kimura:03}

However, within the scenario of partially frustrated Mn-spins the
role of the magnetic $A$-sites is not completely clear so far.
Generally, the moments of the rare-earth ions are polarized due to
the coupling with the Mn subsystem resulting in a noticeable
anisotropic contribution to the low-temperature magnetic and
thermodynamic properties of the manganites as e.g.\ recently
studied in detail for the systems \PMO\ and
\NMO.\cite{hemberger:02} As analyzed for the case of TbMnO$_3$ by
Quezel \etal\ \cite{quezel:77,kenzelmann:05} the overall magnetic
structure in the rare-earth manganites can be rather complex. They
found a sine-wave ordering of the Mn$^{3+}$ moments with the
ordering wave vector along the $b$-axis below 40~K and a
short-range incommensurate ordering of the Tb$^{3+}$ moments with
a different wave vector below 7~K. At the ordering temperature of
the rare-earth moments significant anomalies in the dielectric
constant can be found denoting the possible influence of the
magnetic $A$-sites. \cite{kimura:05}

In this paper we report structural, magnetic susceptibility,
magnetization, and specific heat measurements for single and
poly-crystalline \EuY. Our focus is directed towards the isovalent
doping of the trivalent $A$-site in $R$MnO$_3$, with $R$ denoting
elements Eu$^{3+}$ (4f$^8$) or Y$^{3+}$
([Kr]).\cite{ivanov:06pssb} This allows for a systematic variation
of the ionic radii and possibly of the Mn-O-Mn angle, which we
relate to the development of the complex magnetic groundstates and
ferroelectric phases analogous to the pure rare-earth compounds
$R$MnO$_3$ with $R=$~Gd, Dy, Tb. The system (Eu:Y)MnO$_3$ offers
the possibility to {\em continuously} control the $A$-site volume
of the orbitally ordered perovskite structure and thus to tune the
corresponding multiferroic phases without the additional influence
of a magnetic rare earth moment. A further aspect of the system
\EuY\ is the robustness of long-range polar
order 
against the influence of purely structural $A$-site disorder.


\section{Experimental Details}

\EuY\ single crystals were grown in Ar flow by a floating-zone
method with radiation heating for Y-concentrations $x = 0, 0.1,
0.2, 0.3,$ and 0.5. Additional concentrations have been prepared
as poly-crystals for higher concentrations up to $x=0.95$ using
conventional solid state reaction methods. Powder-diffraction
experiments were performed on powder of crushed single and
poly-crystals at room temperature with a STOE diffractometer
utilizing Cu-K$_{\alpha}$ radiation with a wave length $\lambda =
0.1541$~nm. The magnetic susceptibility and the magnetization were
recorded using a commercial SQUID magnetometer for temperatures $T
<$~400~K and external magnetic fields up to 50~kOe. The dielectric
constant has been measured employing a frequency response analyzer
({\sc Novocontrol $\alpha$-analyzer} and {\sc HP4284A }) and the
spontaneous electric polarization was recorded as integrated
pyro-current.
The specific heat was measured in a PPMS-system ({\sc
Quantum-Design}).

\section{Results and discussion}

\subsection{Structure}

\begin{figure}[tb]
\includegraphics[clip,width=0.8\columnwidth]{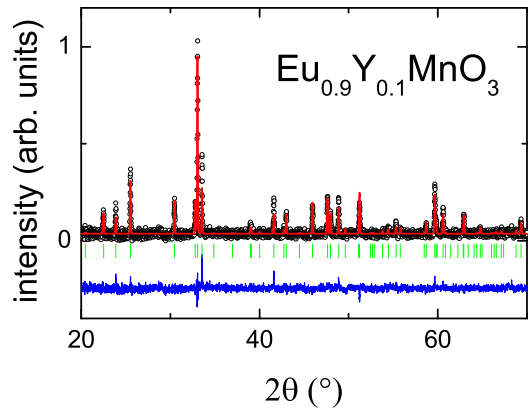}
\caption{(Color online) Representative powder XRD-Pattern for the
Eu$_{0.9}$Y$_{0.1}$MnO$_3$. The lines show the results of a
Rietveldt refinement (upper line through the data points) and the
corresponding deviation from the experimental data (below the
data). The bars in between denote reflections of the Pbnm
structure.
\label{xrd}}
\end{figure}

\begin{figure}[tb]
\includegraphics[clip,width=0.95\columnwidth]{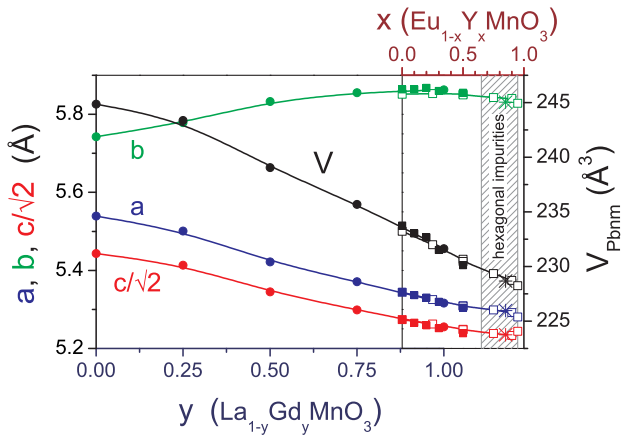}
\caption{(Color online) Lattice constants $a$, $b$, and
$c/\sqrt{2}$ (left scale) and volume of the unit cell (right
scale)
in \EuY\ and \LaGd\ vs.\ concentration of Y ($x$, upper scale,
squares) and Gd ($y$, lower scale, circles). Both scales are
shifted to adapt the overlap of the data. The solid symbols
represent data evaluated from regrinded single crystals, the open
symbols represent data evaluated from polycrystalline material.
The asterisks denote the lattice parameters of pure TbMnO$_3$
which are placed to match the values of \EuY\ close to
$x\approx0.85$. Within the hatched area small traces of hexagonal
impurity phases ($<2\%$) have been detected.
\label{abcv}}
\end{figure}

A typical result for the structural characterization of the
samples by x-ray diffraction is shown in Fig.~\ref{xrd}. The
diffraction patterns were refined using Rietveld analysis. All
samples investigated revealed the O' orthorhombic structure
(Pbnm). No impurity phases were detected above background level
for Y-concentrations $x<0.75$. For higher concentrations (as
denoted in Fig.~\ref{abcv}), small traces of the hexagonal (P63cm)
phase of \YMO\ show up which could be estimated to be below 2~\%
and will not considerably influence the regarded macroscopic
properties. The lattice constants and the volume of the unit cell
as derived from the profile analysis are shown in Fig.~\ref{abcv}.
In addition, the data for the system \LaGd, which was taken from
Ref.~[\onlinecite{hemberger:04prb}], is displayed on a separate
scale. Both scales, for Gd-doping $y$ and Y-doping $x$ are shifted
to achieve an overlap of the structural data. For all
concentrations we find $b
> a > c/\sqrt{2}$ indicative for a static JT distortion superimposed
on the high temperature O-type (i.e.\ not JT distorted)
orthorhombic structure, which results from the buckling and
tilting of the MnO$_6$ octahedra due to geometrical constraints.
However, for \EuY, despite the continuous shrinking of the lattice
volume $V$, the orthorhombic distortion parameterized by
$\varepsilon = (b - a)/(a + b)$ tends
to saturate 
for higher Y-concentrations. The inequality of the lattice
constants $a$ and $b$ reflects the tilting of the octahedra around
the $b$-axis and implies significant deviations from $180^\circ$
of the Mn-O-Mn bond angle $\phi$ within the $ab$-plane. Another
quantity reflecting the development of $\phi$ is the
tilting distortion along $b$ 
of the $A$-site as obtained from the Rietveld refinement of the
atomic positions. This parameter (not shown) is almost constant
for the system \EuY\ in contrast to the less distorted system
\LaGd.\cite{hemberger:04prb} This implies, that the alteration of
the spin system due to the influence of the Mn-O-Mn bond angle may
not be the only valid mechanism for \EuY. In addition, the
variation of the magnetic exchange via the reduction of the volume
and the corresponding changes in the orbital overlap may also have
to be considered. Also the $A$-site disorder and the corresponding
variance in the $A$-site ionic radii could play an important role.
The asterisks in Fig.~\ref{abcv} display the structural parameters
of pure \TMO.\cite{quezel:77} The data agree with the findings for
the Y-concentration $x\approx0.85$. If only taking into account
the averaged ionic radii, the values of the Tb-system should be
located at $x\approx0.4$.\cite{noda:06} For the system \GMO\ an
equivalent concentration of $x\approx0.3$ is found while the value
is expected to be $x\approx0.15$. This shows that for \EuY\ the
lattice is less contracted than it would be expected compared to
the pure rare earth manganites.

\subsection{Magnetism}

\begin{figure}[tb]
\includegraphics[clip,width=0.9\columnwidth]{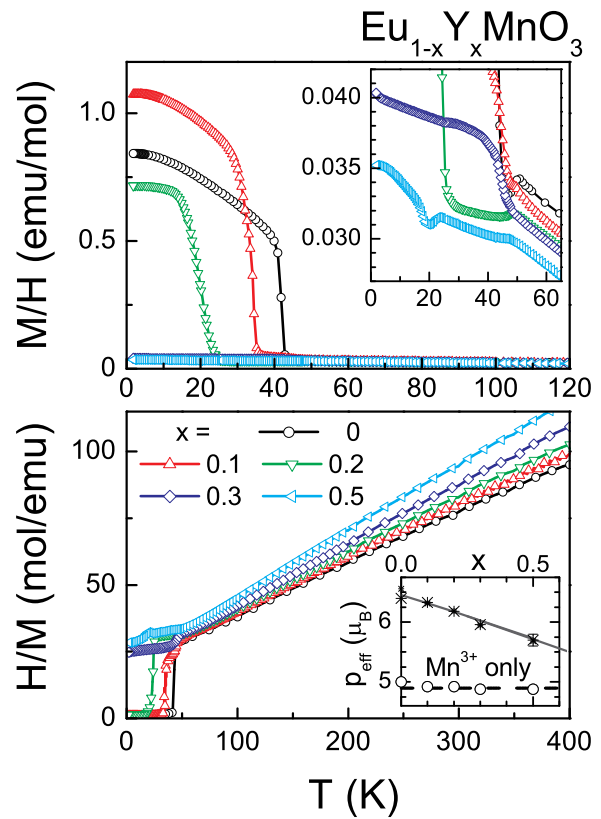}
\caption{(Color online) Temperature dependence of the magnetic
DC-susceptibility $\chi=M/H$ (upper frame, for $T<100$~K) and the
inverse DC-susceptibility $1/\chi=H/M$ (lower frame, for
$T<400$~K) for various concentrations of \EuY\ as measured along
$c$ in an external magnetic field of 1~kOe. The upper inset
proveides an expanded view of $M/H$ close to the magnetic phase
transitions. The lower inset shows the effective paramagnetic
moment $p_{\rm eff}$ as obtained from a simple Curie-Weiss type
evaluation ($\ast$) of the linear regime of $1/\chi$ above
$T=100$~K and from an analysis separating the contributions of Eu
and Mn as described in the text ($\circ$).
\label{chi}}
\end{figure}

\begin{figure}[tb]
\includegraphics[clip,width=0.8\columnwidth]{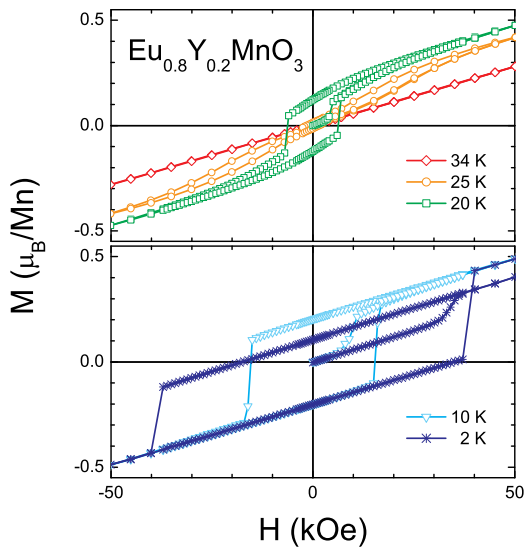}
\caption{(Color online) Magnetization along $c$ of
Eu$_{0.8}$Y$_{0.2}$MnO$_3$ as function of the external magnetic
field for a series of temperatures between 2~K and 34~K.
\label{maghys}}
\end{figure}

Fig.~\ref{chi} shows the magnetic DC susceptibilities $\chi=M/H$
for concentrations $x \leq 0.5$. At elevated temperatures all
susceptibilities follow a Curie-Weiss (CW) law as revealed by the
inverse representation in the lower frame of Fig.~\ref{chi}. The
corresponding effective paramagnetic moments $p_{\rm eff}$ are
plotted in the lower inset of Fig.~\ref{chi} ($\ast$). $p_{\rm
eff}$ is decreasing due to the decreasing rare-earth contribution.
Y$^{3+}$ is nonmagnetic and for Eu$^{3+}$ the $4f^6$-configuration
leads to $J = 0$. However, for Eu low-lying multiplets give rise
to a Van-Vleck type of contribution, which in this temperature
regime is nearly Curie-like and thus enhances the effective
paramagnetic moment.
For a more detailed analysis the paramagnetic high temperature
behavior was fitted by a sum of the contribution of Eu$^{3+}$
($\propto(1-x)$) and the CW type contribution of Mn$^{3+}$. For
the Eu-contribution we used the well known free ion
susceptibility of Eu$^{3+}$ determined by 
its excited multiplets with $E_1 \approx 500$~K and $E_2\approx
3E_1$ for $J=1$ and $J=2$, respectively.\cite{taylor:72,herpin:68}
At high temperatures it is linear in $1/T$ while at low $T$ it is
determined by a temperature independent Van-Vleck contribution.
All our data can be nicely described using $p_{\rm eff,Mn}$ as
plotted in the lower inset of Fig.~\ref{chi} ($\circ$). The
manganese contribution turns out to be independent of $x$ and
agrees well with the expected theoretical value
$2\mu_B\sqrt{[S(S+1)]} = 4.9 \mu_B$.
For concentrations $x\leq0.5$ the CW-temperatures of the
Mn-subsystem are $\Theta \approx - 58 \pm 4$~K and are almost
independent of $x$. This result is in agreement with corresponding
findings in \LaGd\ for high Gd
concentrations.\cite{hemberger:04prb}

Below 50~K all compositions undergo a magnetic transition into an
incommensurate antiferromagnetic (IC) phase as concluded in
analogy to the pure rare earth compounds.\cite{goto:04} This
transition was clearly detected by heat-capacity experiments,
which will be discussed later. It can only be seen as a small
change of slope in the temperature dependence of the magnetic
susceptibility as it is not connected with the onset of a
ferromagnetic component (see upper inset of Fig.~\ref{chi}). A
much stronger anomaly appears at the transition from the
incommensurate magnetic state into the canted A-type
antiferromagnetic (CAFM), which takes place for concentrations $x
\leq 0.3$. As revealed by the upper frame of Fig.~\ref{chi}, the
onset of CAFM order at $T_{\rm CAFM}$ shows up as a strong upturn
of magnetization indicating a significant ferromagnetic
contribution. This FM component arises due to spin canting, which
can be explained by the Dzyaloshinsky-Moriya interaction and is an
intrinsic feature of the CAFM state in $R$MnO$_3$
systems.\cite{solovyev:96,skumryev:99} As stated above, $T_{\rm
CAFM}$ continuously decreases from 140~K in LaMnO$_3$ down to
about 20~K in GdMnO$_3$. For the system \EuY\ this transition is
shifted to $T\approx0$ for Y-concentrations of about $0.2\leq x
\leq 0.3$. For \TMO\ and for concentrations $x \geq 0.3$ in \EuY\
no A-type AFM can be found but the character of the magnetic phase
changes from a sinusoidal to a helicoidal (spiral) magnetic
structure at temperatures between 20 and 30~K.\cite{kenzelmann:05}
This latter transition is connected with the onset of
ferroelectric order as will be discussed later. However, it still
seems a controversial debate if this spiral AFM structure remains
incommensurate \cite{kenzelmann:05} or locks-in at a commensurate
wave vector. \cite{arima:05} It should be noted that at least for
the composition $x=0.5$ additional small anomalies can be found at
the FE transitions at 24~K and 18~K, indicative for a partial
rearrangement of the magnetic structure.

Fig.~\ref{maghys} shows the magnetization $M(H)$ for
Eu$_{0.8}$Y$_{0.2}$MnO$_3$ in magnetic fields up to 5~T as
measured after zero-field cooling. At $T=34$~K (upper frame of
Fig.~\ref{maghys}) an apparently paramagnetic magnetization
behavior is detected. As can be deduced from the specific heat
data (presented below) the magnetic state of the sample at this
temperature is in fact the IC phase. At 25~K the CAFM state can be
induced by the magnetic field as reflected by the twofold
hysteresis loop with the typical signature of a smeared out
metamagnetic phase transition at $H\approx$~30~kOe. On further
decreasing temperature this metamagnetic transition becomes fully
irreversible and the two loops are merged into one as it is
characteristic for a weak ferromagnet. After the initial increase
of the field the CAFM state persists and only the ferromagnetic
component is switched at a sharply defined coercive field, which
increases up to $H_c\approx$~40~kOe for the lowest temperatures.
At 2~K (lower frame of Fig.~\ref{maghys}) even an intermediate
state with a smaller ferromagnetic component is stabilized after
the initial increase of the magnetic field leading to an
asymmetric shape of the hysteresis loop.

\subsection{Dielectric properties}

\begin{figure}[b]
\includegraphics[clip,width=0.95\columnwidth]{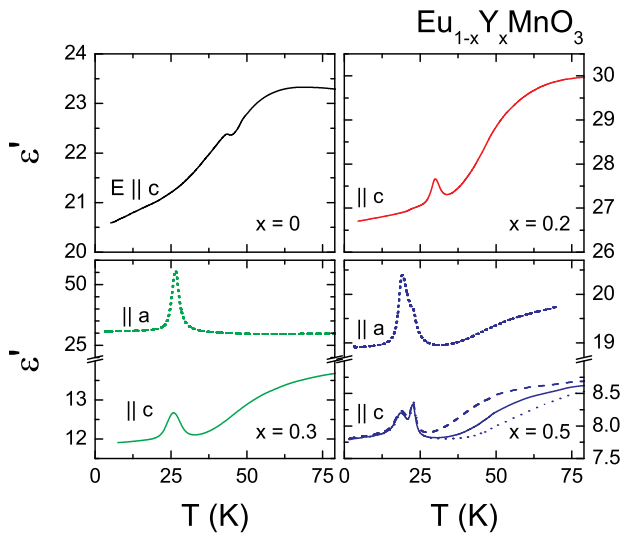}
\caption{(Color online) Temperature dependence of the real part of
the permittivity $\varepsilon'$ for the Y-concentrations $x=0$,
0.2, 0.3, and 0.5 as measured at $\nu = 100$~kHz. The lower right
frame ($x=0.5$) shows additional data for the frequencies 27~kHz
(dashed) and 250~kHz (dotted). Additional data for $E \| a$ is
given for the concentrations $x=0.3$ and 0.5 (lower frames, short
dashes). In general, the measured data of $\varepsilon'$ may
comprise a constant additional contribution due to stray
capacitance of the measurement setup together with an relative
uncertainty of up to 25~\% due to the determination of the sample
geometry. Hence, the absolute values of $\varepsilon'(T)$ are not
discussed. However, the relative accuracy of the measurement is
much higher and lies within the thickness of the lines.
%
%
\label{dk}}
\end{figure}

\begin{figure}[tb]
\includegraphics[clip,width=0.8\columnwidth]{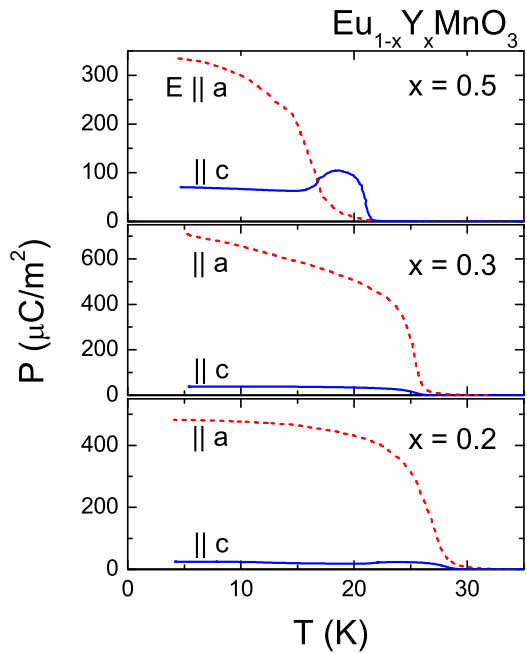}
\caption{(Color online) Electric polarization as measured on
heating after cooling down in a poling field of about
100~V/mm for the Y-concentrations $x=0.2$, 0.3, and 0.5.
%
\label{pyro}}
\end{figure}

Fig.~\ref{dk} displays the real part of the dielectric constant as
measured in the radio-frequency range for various concentrations
of \EuY. All $\varepsilon'(T)$ curves show a steplike increase
with increasing temperature along the $c$-direction. Its point of
inflection is located around 50~K but as shown exemplarily for the
concentration $x=0.5$ (lower right frame of Fig.~\ref{dk}) its
position is strongly frequency dependent. Similar phenomena have
already been found e.g.\ for \TMO\ and have been ascribed to the
relaxation of localized polarons.\cite{kimura:03} In the following
we will concentrate on the anomalies below 45~K. For pure \EMO\
only a tiny kink can be found near $T\approx 43$~K. This coincides
with the onset of weak ferromagnetism in the CAFM phase as
revealed from the magnetic susceptibility measurements. For
$x=0.2$ a similar anomaly at $T_{\rm CAFM}\approx 22$~K can be
observed, which can hardly be recognized, but a much more
prominent peak occurs at 30~K. For $x=0.3$ this anomaly is located
at 26~K and for $x=0.5$ it forms a double-peak structure at
roughly 23~K and 19~K. A similar double structured peak can also
be detected along the $a$-direction, however with shifted weight
from higher to the lower temperature peak. These anomalies in
$\varepsilon'(T)$ coincide with the onset of ferroelectricity as
will be described later. For the concentrations $x=0.2$, 0.3, and
0.5 the electric polarization $P(T)$ is shown in Fig.~\ref{pyro}.
As revealed by the uppermost frame of Fig.~\ref{pyro}, obviously
the double structure in $\varepsilon'(T)$ for
Eu$_{0.5}$Y$_{0.5}$MnO$_3$ (lower right frame of Fig.~\ref{dk})
corresponds to the partial spontaneous reorientation of $P$ from
the $c$- to the $a$-direction. This is in accord with recent
results obtained for $x\approx0.4$ by Kuwahara and
coworkers.\cite{noda:06} Such a spontaneous reorientation of the
electric polarization is not reported for the pure systems
$R$MnO$_3$. \cite{kimura:05} For the compositions \EuY\ with
$x=0.3$ and 0.2 no temperature dependent reorientation of the
polarization could be observed as it is shown in the lower frames
of Fig.~\ref{pyro}. It is remarkable that despite the
magnetoelectric origin of the FE order no clear anomaly in $P(T)$
can be observed for $x=0.2$ around 23~K where a spontaneous FM
component sets in. However, as can be seen from Fig.~\ref{maghys}
the values for the spontaneous magnetization stay below $0.2
\mu_B$ per Mn$^3+$ ion. This is only 1/20th of the maximal ordered
moment of $4 \mu_B$. Thus the effective FM moment can be explained
by a canting of the presumably spiral magnetic structure of less
than $3^\circ$, which represents an only small modification of the
magnetic structure.

Ferroelectricity (FE) does not occur for concentrations below
$x<0.2$. Nevertheless, it is remarkable that a canted, i.e.\
weakly ferromagnetic state and FE are not mutually exclusive. For
$x=0.2$ the system first becomes ferroelectric at $T_{\rm
FE}\approx 30$~K and then the ferromagnetism evolves roughly 8~K
below. In a strict sense, this is the only multiferroic system
among the rare earth perovskite manganites where the ground state
seems to combine the properties of ferromagnetism and
ferroelectricity. The most prominent examples like \TMO\ never
show a finite spontaneous FM and FE order-parameter at the same
time.

\subsection{Specific heat}

\begin{figure}[b]
\includegraphics[clip,width=0.9\columnwidth]{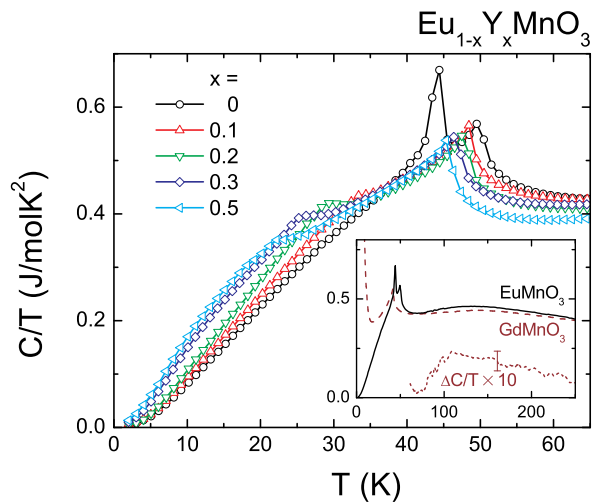}
\caption{(Color online) Heat capacity of \EuY\ plotted as $C/T$ as
a function of temperature.  The inset compares the data for pure
\EMO\ (solid line) and \GMO\ (dashed line) for temperatures up to
250~K. The dotted line represents the difference multiplied by a
factor of 10.
\label{heat}}
\end{figure}

Fig.\ \ref{heat} presents the results of heat-capacity
measurements in \EuY\ as function of temperature. The heat
capacity, plotted as $C/T$, is shown for temperatures $T < 65$~K
and concentrations $0\leq x \leq0.5$. A well defined lambda-like
anomaly shows up between 45~K and 50~K for all concentrations
under investigation. In agreement with magnetization data and
deduced from observations in other rare-earth manganites this
anomaly has to be interpreted as onset of the IC spin structure,
which gives no macroscopic FM moment. \cite{kimura:03} The
transition temperature is slightly decreasing with increasing
Y-concentration. Below 45~K for each concentration a further
anomaly appears at lower temperatures which for pure \EMO\ is
characterized by a sharp peak in $C/T$ at 44~K corresponding to
the transition into the A-type AFM-phase (see Fig.~\ref{chi}). For
the Y doped samples this feature gets considerably smeared out
presumably due to the influence of disorder. Again the
corresponding transition temperature as denoted by the onset of
the broad shoulder-like contribution is monotonously decreasing
with increasing $x$ down to about 20~K for $x=0.5$. However, even
though the shape of this anomaly in $C/T$
is not altered qualitatively, the corresponding transition refers
to different ground states for different compositions. As it was
already revealed in Figs.~\ref{chi} and \ref{pyro} the onset of
spontaneous magnetization corresponds to the transition into a new
magnetic ground state with FM and FE components for concentrations
$x \approx 0.2$. For higher concentrations the spiral IC structure
persists down to lowest temperatures in analogy to the pure rare
earth compounds, but the onset of a FE polarization corresponds to
a change from a sinusoidal to a helicoidal magnetic
structure.\cite{kenzelmann:05} The absence of any qualitative
differences in the curvature of $C/T$ for the different
compositions reflects that the corresponding ground states
together with their partially frustrated spin structures are
nearly degenerate.

It is worth mentioning that the low temperature specific heat in
this system differs essentially from that of \GMO\ or \TMO\ due to
the missing contribution of the magnetic rare
earths.\cite{kimura:03,hemberger:04prb}
Neither a Schottky-type contribution as found due to the rare
earth spins, which are polarized in the effective exchange field
from the CAFM Mn-sublattice, nor an anomaly due to the subsequent
ordering of the magnetic rare earth sublattice in the temperature
range below 10~K can be detected for \EuY. For comparison the
inset of Fig.~\ref{heat} shows data for pure \GMO\ and \EMO.
Besides the mentioned differences at low temperatures for \EMO\ an
additional contribution at higher temperatures is observed. This
contribution can not be explained by the slight differences of the
phonon spectrum because Gd and Eu have very similar masses.
$\Delta C(t)$ shows a broad peak at around 120~K and can be
explained by the relatively small splitting of the Eu-multiplet,
which also gives rise to the Van-Vleck contribution to the
magnetic susceptibility, discussed above.

\section{Phase diagram}

\begin{figure}[tb]
\includegraphics[clip,angle=-0,width=0.8\columnwidth]{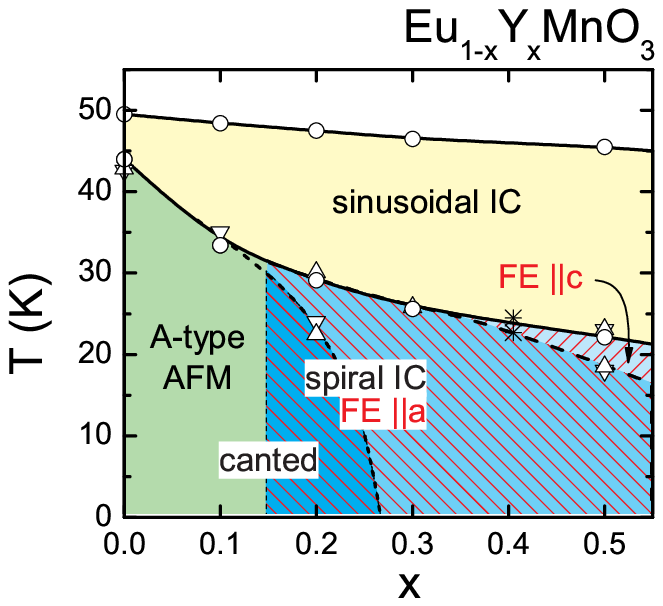}
\caption{(Color online) $(T,x)$-phase diagram of \EuY\ for
Y-concentrations $0\leq x\leq0.5$. The data points were obtained
from the measurements of the specific heat ($\circ$), the
magnetization ($\triangledown$), the permittivity ($\triangle$),
and from Ref.~[\onlinecite{noda:06}] ($\ast$). \label{phase}}
\end{figure}

Based on the results shown above, from the magnetic and dielectric
measurements as well as from the heat capacity, a detailed
$(x,T)$-phase diagram can be constructed and is shown in
Fig.~\ref{phase}. The paramagnetic regime above $T_N \approx
45-50$~K, is followed by the sinusoidal IC structure of the
manganese moments, an interpretation guided by results for
TbMnO$_{3}$ \cite{quezel:77,kenzelmann:05} and o-HoMnO$_{3}$.
\cite{brinks:01} For small Y-concentrations, $x\leq0.2$, this
incommensurate spin-structure locks into the weakly ferromagnetic
A-type CAFM phase. For higher concentrations, $x\geq0.3$, the
ground state is ferroelectric but no ferromagnetic component can
be found. Obviously, the spin-structure does not lock into the
CAFM anymore. However, like in \TMO\ it remains unclear so far, if
the magnetic structure still is incommensurate
\cite{kenzelmann:05} or if a commensurate long wavelength magnetic
vector is established.\cite{arima:05} In a small regime around
$x\approx 0.2$ weak ferromagnetism (CAFM) and weak
ferroelectricity coexist in contrast to the pure multiferroic rare
earth manganites. Regarding the volume of the unit cell, the
series \EuY\ covers the values of \GMO\ and \TMO. Pure \EMO\ is
not ferroelectric. \GMO\ is ferroelectric along $a$ in a weak
external magnetic field along $b$, which destroys the A-type AFM
structure.\cite{kimura:05} For \EuY\ a concentration slightly
lower than $x=0.2$ probably will reproduce this property of
GdMnO$_3$. However, according to Fig.~\ref{abcv} the Eu:Y-system
corresponding to \GMO\ should have $x\approx 0.2$. This
composition is already ferroelectric without magnetic field even
though a canted AFM structure persists.
Pure \TMO\ is ferroelectric along $c$.\cite{kimura:05} However,
concerning the volume of the unit cell it could not be reproduced
by pure single crystalline \EuY\ close to $x\approx 0.85$ as shown
in Fig.~\ref{abcv}. In \TMO\ there is an additional anisotropy due
to Tb-ions and Tb-Mn exchange which could lead to the observed
electric polarization along the $c$-axis. This should be taken
into account for the comparison between \TMO\ and \EuY\ systems.
However, the tendency towards a spontaneous ferroelectric
component along $c$ is already indicated also for \EuY\ at lower
Y-concentrations like $x=0.5$ (Fig.~\ref{pyro}).

It is interesting that only the weak FE component along $a$
coexists with spontaneous weak ferromagnetism, namely within the
concentration range around $x\approx 0.2$, while for higher
concentrations, where at least an intermediate FE component along
$c$ is detected, no weak ferromagnetism shows up. According to the
symmetry properties of the magnetoelectric coupling for the rare
earth manganites\cite{kenzelmann:05,mostovoy:06,katsura:05}, the
ferroelectric polarization vector is orthogonal to both, the
vector of the magnetic spiral and the magnetic modulation vector,
which points along the $b$ axis.\cite{goto:04} For \TMO
\cite{kenzelmann:05} and analog for the higher Y-concentrations of
\EuY\ the ferroelectric component points along the $c$-direction,
which is compatible with a spiral vector along $a$ (i.e.\ the
spins rotating in the $bc$-plane). For \GMO\ (in small magnetic
fields along $c$\cite{baier:06}) and similarly for the
concentrations around $x\approx 0.2$ of \EuY, the ferroelectric
component points along the $a$-direction. This would be compatible
with a spiral vector along $c$ (i.e.\ the spins rotating in the
$ab$-plane). The homogeneous canting of this helicoidal spin
structure out of the $ab$-plane could then generate the observed
ferromagnetic moment along $c$.

\section{Conclusion}

In summary, we have characterized poly- and single-crystalline
samples of the system \EuY\ by means of structural, magnetic,
thermodynamic, and dielectric measurements. In comparison with
recently published results for the pure rare earth systems we
constructed a $(T,x)$-phase diagram for compositions $x \leq 0.5$.
The ground state \EuY\ changes from a canted A-type AFM without long
range polar order for concentrations $x<0.15$ towards a presumably
incommensurate spiral magnetic structure coexisting with a
ferroelectric component for $x\geq 0.3$. For these higher
Y-concentrations the orientation of the ferroelectric component
changes spontaneously from the $c$-axis at higher temperatures
towards the $a$-axis for low temperatures. The regime with  $P
\parallel c$ increases with increasing Y-concentration. In this
sense \EuY\ resembles the development of the ferroelectric
orientation with decreasing ionic $A$-site radius as in the system
\TbGd. \cite{goto:05} Hence, all essential magnetoelectric
properties existing in the pure rare earth compounds \GMO\ and
\TMO\ are reproduced. The influence of the magnetic rare earth
seems to be restricted to the distortion of the magnetic structure
connected with the magnetic $A$-site ordering at low temperatures.
The modulation of the magnetic structure related to the onset of
ferroelectricity is realized due to the Mn sublattice only.
However, the contraction of the lattice with increasing $x$ in
\EuY\ is less effective than estimated from the average ionic
$A$-site radii in comparison to the pure rare earth compounds,
which points towards the importance of disorder effects. The
latter may also influence the properties in the concentration
range around $x\approx0.2$ where a finite spontaneous magnetic
moment and ferroelectricity coexist. The details of the underlying
magnetic structure are not yet verified but it is plausible to
expect a spontaneously canted spiral. Such a FM and FE ground
state has not been found in the pure rare earth compounds so far
and deserves further study.


\acknowledgments

This work was partly supported by the Bundesministerium f\"ur
Bildung und Forschung (BMBF) via Grant No. VDI/EKM 13N6917-A, by
the Deutsche Forschungsgemeinschaft via  Sonderforschungsbereich
SFB 484 (Augsburg), and by INTAS and Russian Foundation for Basic
Researches N 04-02-16592, 06-02-17514, 04-02-81046-Bel2004.

\bibliographystyle{apsrev}
\bibliography{jabbr,joalt,joneu}



\end{document}